\documentclass[aps,prd,superscriptaddress,twocolumn,longbibliography]{revtex4-2}

\usepackage{graphicx}	
\graphicspath{{./}{figures/}{figures/skymap_coverage/}{figures/roc/}}
\usepackage{amsmath}	
\usepackage{amssymb}	
\usepackage{subfigure}
\usepackage{bbm}
\usepackage{siunitx}
\usepackage{orcidlink}

\begin{document}


\title{Using overlap of sky localization probability maps for filtering potentially lensed pairs of gravitational-wave signals}

\thanks{Shared first authorship: Henry W. Y. Wong, Lok W. L. Chan}


\author{Henry~W.~Y.~Wong\,\orcidlink{0000-0002-4027-9160}}
\email[]{Contact author: wyhwong@link.cuhk.edu.hk}
\affiliation{Department of Physics, The Chinese University of Hong Kong, Shatin, New Territories, Hong Kong}
\affiliation{Department of Statistics and Data Science, University of Texas at Austin, Austin, Texas, Texas 78712, USA}

\author{Lok~W.~L.~Chan\,\orcidlink{0009-0006-1431-1784}}
\email[]{lokwlc@link.cuhk.edu.hk}
\affiliation{Department of Physics, The Chinese University of Hong Kong, Shatin, New Territories, Hong Kong}

\author{Isaac~C.~F.~Wong\,\orcidlink{0000-0003-2166-0027}}
\email[]{chunfung.wong@kuleuven.be}
\affiliation{KU Leuven, Department of Electrical Engineering (ESAT), STADIUS Center for Dynamical Systems, Signal Processing and Data Analytics, Kasteelpark Arenberg 10, 3001 Leuven, Belgium}
\affiliation{Leuven Gravity Institute, KU Leuven, Celestijnenlaan 200D box 2415, 3001 Leuven, Belgium}
\affiliation{Department of Physics, The Chinese University of Hong Kong, Shatin, New Territories, Hong Kong}

\author{Rico~K.~L.~Lo\,\orcidlink{0000-0003-1561-6716}}
\email[]{kalok.lo@nbi.ku.dk}
\affiliation{Center of Gravity, Niels Bohr Institute, Blegdamsvej 17, 2100 Copenhagen, Denmark}
\affiliation{LIGO Laboratory, California Institute of Technology, Pasadena, California, California 91125, USA}

\author{Tjonnie~G.~F.~Li\,\orcidlink{0000-0003-4297-7365}}
\email[]{tjonnie.li@kuleuven.be}
\affiliation{Department of Physics, The Chinese University of Hong Kong, Shatin, New Territories, Hong Kong}
\affiliation{Institute for Theoretical Physics, KU Leuven, Celestijnenlaan 200D, B-3001 Leuven, Belgium}
\affiliation{KU Leuven, Department of Electrical Engineering (ESAT), STADIUS Center for Dynamical Systems, Signal Processing and Data Analytics, Kasteelpark Arenberg 10, 3001 Leuven, Belgium}
\affiliation{Leuven Gravity Institute, KU Leuven, Celestijnenlaan 200D box 2415, 3001 Leuven, Belgium}


\date{\today}

\begin{abstract}

	Strong gravitational lensing creates multiple images of a gravitational wave transient.
	The current state-of-the-art method for identifying such lensing events is a computationally expensive full Bayesian analysis.
	In this paper, we investigate the feasibility and efficiency of using the overlap of sky localization probability maps (skymaps) to quickly filter potentially lensed gravitational wave signal pairs.
	We introduce three overlap statistics and test their performance using 200 simulated lensed pairs of gravitational-wave signals across five sets of signal-to-noise ratios.
	By setting a threshold with a false positive rate of $\mathrm{FPR} = 10^{-2}$ for the three overlap statistics, we find that we can filter out over $99\%$ of nonlensed events while retaining all lensed events.
	The statistics for each event pair can be computed instantly, and can be used in practice to quickly analyze existing events using the skymaps from the low-latency localization pipelines when results from the full parameter estimation are not available.

\end{abstract}


\maketitle

\section{Introduction}
\label{sec:introduction}

When a gravitational wave (GW) passes by a massive object, such as a black hole or a galaxy cluster, its path is bent toward the massive object~\cite{1010.3829}, a phenomenon known as gravitational lensing.
As a result, the GW can take different paths to reach the observer and form multiple images with different arrival times~\cite{1974IJTP....9..425O,1986PhRvD..34.1708D,astro-ph/9605140,1998PhRvL..80.1138N,astro-ph/0305055}.
Consequently, it is predicted that a number of GW signals with different amplitudes and arrival times originating from the same source could be observed.

Since the detection of the first GW signal in 2015~\cite{1602.03837}, 93 confident GW events have been detected by Advanced LIGO~\cite{1411.4547} and Advanced Virgo~\cite{1408.3978} in the first (O1), the second (O2), and the third observing run (O3)~\cite{1811.12907,2010.14527,2108.01045,2111.03606}.
Although there is no compelling evidence in lensing of the observed GW events in O3a~\cite{2105.06384}, theoretical predictions suggest roughly one strong GW lensing event per year at the Advanced LIGO-Advanced Virgo-KAGRA design sensitivity~\cite{2106.06303,2105.14390}.
Therefore, it is expected that the first lensed GW event pair will be detected in the coming years~\cite{1703.06319,2106.06303}.

There are different methods to detect strongly lensed GW event pairs~\cite{2105.06384}.
One method~\cite{1807.07062,1901.02674} uses the overlap of the posterior distribution of the source parameters to check for the consistency of the source properties between events.
The overlap of the two posterior distributions is expected to be high when the two events come from the same source.
Another method~\cite{2009.06539,2104.09339} performs a full Bayesian analysis, namely the joint-parameter estimation (joint-PE), on the detected events.

Joint-PE provides a solid statistical framework to calculate the probability of a GW event pair being lensed images from the same source.
The current number of confident GW events is 93~\cite{1811.12907,2010.14527,2108.01045,2111.03606}, which gives 4278 pairs of GW events.
However, performing joint-PE on all event pairs is very time consuming.
Moreover, for a proposed Einstein Telescope survey, the predicted event rate of binary neutron star mergers is as high as $O(10^5 - 10^6)$ $\mathrm{yr}^{-1}$ and $O(10^4)$ $\mathrm{yr}^{-1}$ for neutron star-black hole mergers~\cite{amaro2009einstein}.
This means that the detector sensitivity and the number of GW signals in the future will be far higher and will further increase the time for doing joint-PE on all possible event pairs. For the purpose of sustainability, it is necessary to establish a low latency filter to screen out event pairs that are definitely not lensed.

There are several fast algorithms for rapid identification of strongly lensed GW events developed to address the sustainability issue.
One of the algorithms is aimed to search for strongly lensed multiple GW images---from the fact that the signals come from the same source, it is possible to split the joint PE into two easier PEs, that we can use the result from one event to perform the inference of the another event~\cite{2105.04536}.
Such methodology enables relatively quick multiple-image analyses with high accuracy.
Another approach proposed by Goyal \textit{et al.} uses a machine learning model to predict the probability of lensing in event pairs from the time-frequency maps and posterior distributions of GW events~\cite{2106.12466}.
The computational time of the lensing probability is estimated to be around three seconds with most of the time spent on loading necessary files.
Both of the mentioned works have successfully reduced the computational expense of filtering potentially lensed GW event pairs.

To establish a low latency filter, we propose the use of sky localization probability maps (skymaps) to analyze the possibility of an event pair coming from the same source.
The deflection angle is usually a few hundred arc seconds for black holes and galaxy clusters~\cite{1992grle.book.....S}, which is negligibly small as it is far less than the uncertainty of the localization of the source~\cite{2016LRR....19....1A}.
Therefore, the skymaps of the images from the same source are expected to be highly overlapped.
In other words, event pairs with nonoverlapping skymaps are definitely not lensed pairs.
We use this fact to analyze the possibility of whether two GW signals are a lensed pair and filter potentially lensed pairs by comparing their skymaps.
With the use of overlap, we can reject a large fraction of event pairs that are definitely not image pairs.
However, one must note that having a high overlap in the skymaps does not necessarily imply the events are a lensed pair. In contrast to the machine learning approach in~\cite{2106.12466}, we investigate using measures of the overlap of skymaps without involving sophisticated learning models, which can be easily incorporated into search pipelines such as the subthreshold search~\cite{1904.06020} to send alerts for performing follow-up analyses.
The goal of the low latency filter is to reduce time cost by rapidly filtering out most of the nonlensed GW event pairs for further lensing analysis.
Therefore, we would like to quantitatively investigate the efficiency of using skymap overlap to filter the nonlensed pairs while not rejecting the lensed pairs.
With a low latency skymap generated by a rapid sky localization algorithm such as \texttt{BAYESTAR}~\cite{1508.03634}, a quick preliminary lensing analysis on two GW signals can be done within a minute.

In this paper, we present three overlap statistics: normalized posterior overlap, 90\% credible region overlap and cross-highest posterior density statistic, to quantitatively investigate the efficiency of using skymap overlap to filter the nonlensed pairs while not rejecting the lensed pairs. The details of the statistics are described in Sec.~\ref{sec:Overlap_Statistics}.
Using simulated GW events, we study the distribution of skymap overlap of lensed GW event pairs under different signal-to-noise ratios (SNRs).
In order to filter potentially lensed event pairs, a reasonable threshold of skymap overlap that can filter out nonlensed event pairs while keeping all of the lensed event pairs should be set with a specific false positive rate (FPR).
Based on the results, we compare the effectiveness of the three statistics and discuss the feasibility of using skymap overlap to filter potentially lensed pairs of GW signals.

This paper is organized as follows: In Secs.~\ref{sec:skymap} and  \ref{sec:Overlap_Statistics}, we establish the methodology of the search by introducing posterior distribution and the three statistics mentioned.
In Sec.~\ref{sec:result}, we investigate the feasibility of using skymap overlap for filtering potentially lensed pairs of GW signals with the receiver operating characteristic (ROC) curves.
In Sec.~\ref{sec:conclusion}, we conclude our findings on the feasibility of skymap overlap in this paper and its potential work in the future.

\section{Skymap}
\label{sec:skymap}
\subsection{Posterior distribution}
\label{mtd:PD}
To generate a skymap, we use the observed data to estimate the source location of a GW projected on the celestial sphere under a fixed waveform model.
This process relies on parameter estimation, which allows us to quantify the uncertainty of the GW location on the celestial sphere.
We first define a hypothesis $H$ that the GW signal fits the fixed waveform model. Then for each GW source parameter, we set up a prior probability distribution $p(\theta|H)$, which represents the \textit{a priori} knowledge of the model parameters before observing the data.
Using Bayes' theorem, the prior probability distribution $p(\theta|H)$ and the data $d$ of the GW signals are used to compute the posterior distribution $p(\theta|d,H)$ of the waveform parameters~\cite{1409.7215} defined as
\begin{equation}
	p(\theta|d,H) \equiv \frac{p(\theta|H)p(d|\theta,H)}{p(d|H)}\,.
\end{equation}

\subsection{Sky localization probability map}
\label{mtd:skymap}
A waveform model typically has many parameters that we can denote as $\theta=\{\theta_1,\theta_2,...,\theta_N\}$.
The posterior distribution $p(\theta|\textit{d,H})$ represents the joint posterior distribution of all parameters. To determine the posterior distribution of a specific parameter, we marginalize the nuisance parameters, where the posterior distribution of a specific parameter $p(\theta_1|d,H)$ is obtained by

\begin{equation}
	p(\theta_1|d,H) \equiv \int \mathrm{d}\theta_2...\mathrm{d}\theta_Np(\theta|d,H) \,.
\end{equation}

After obtaining the posterior distribution of the right ascension and the declination of the source location, which is referred to as a skymap in this paper, we compute the three overlap statistics of a skymap pair described in the following Section.

\section{Overlap Statistics}
\label{sec:Overlap_Statistics}
\subsection{Posterior overlap}
\label{mtd:stat_PO}
After performing the parameter estimation on two GW events separately, we obtain two posterior distributions $p(\hat \Omega)$ and $q(\hat \Omega)$ of their skymap coordinate $\hat \Omega$.
We determine whether a GW signal pair is potentially lensed by the inner product of $p(\hat \Omega)$ and $q(\hat \Omega)$, which quantifies the overlap between two probability densities.
The posterior overlap $\mathcal{D}_{\mathrm{PO}}$ between two skymaps is defined similarly to~\cite{1807.07062} as
\begin{equation}
	\mathcal{D}_{\mathrm{PO}} \equiv \int p(\hat \Omega')q(\hat \Omega') \mathrm{d}\Omega'
\end{equation}
which is not normalized.
In order for the posterior overlap of different GW signal pairs to be comparable to each other, we normalize the posterior overlap with skymap of the two GW events, which is defined as
\begin{equation}
	\mathcal{D}_{\mathrm{NPO}} \equiv \frac{\int p(\hat \Omega')q(\hat \Omega') \mathrm{d}\Omega'}{\sqrt{\int p(\hat \Omega')p(\hat \Omega') \mathrm{d}\Omega'} \sqrt{\int q(\hat \Omega')q(\hat \Omega') \mathrm{d}\Omega'}} \in [0,1]\,.
\end{equation}

\subsection{90\% credible region overlap}
\label{mtd:stat_90CR}
For a GW signal with an extremely large SNR, the posterior distribution of the source location could be narrowly peaked.
This may lead to an extremely small value of posterior overlap and normalized posterior overlap with another GW signal.
As a result, the two GW signals may be recognized as a nonlensed pair and filtered out.
Therefore, we propose to use a 90\% credible region overlap.

Instead of finding the overlap probability of the source position on the sky, we directly compute the overlapping area of the 90\% credible regions.
The overlapped area is further normalized by the minimum area between two 90\% credible regions.
The minimum value is used to keep a high 90\% credible region overlap when the difference between two event SNRs is high for lensed pairs.
Therefore, it reduces the chance of lensed pairs being filtered out.
The 90\% credible region overlap $\mathcal{D}_{\mathrm{90\%CR}}(p,q)$ of two GW signals is computed by masking their skymaps, which is defined as \\
\begin{equation}
	\begin{aligned}
		 & \mathcal{D}_{\mathrm{90\%CR}}(p,q)                                                                                                                                                                                                                                                                                                                                                                       \\
		 & =\frac{\displaystyle\int\mathbbm{1}_{\mathrm{90\% CR}} \left[ p(\hat \Omega')\right] \mathbbm{1}_{\mathrm{90\% CR}} \left[ q(\hat \Omega') \right] \mathrm{d}\Omega'}{{\mathrm{min}} \left( \displaystyle \int\mathbbm{1}_{\mathrm{90\% CR}}\left[ p(\hat \Omega') \right] \mathrm{d} \Omega',\displaystyle\int\mathbbm{1}_{\mathrm{90\% CR}}\left[q(\hat \Omega')\right] \mathrm{d}\Omega' \right)} \,,
	\end{aligned}
\end{equation}
where $\mathbbm{1}_{\mathrm{90\% CR}} \left[ p(\hat \Omega)\right]$ and $\mathbbm{1}_{\mathrm{90\% CR}} \left[ q(\hat \Omega)\right]$ are the indicator function. 
This indicator function is 1 when the sky location is in the 90\% credible region of the corresponding skymap, and 0 otherwise.

\subsection{Cross-HPD statistic}
\label{mtd:stat_crossHPD}
We introduce another statistic called the cross-HPD.
The purpose of this statistic is similar to that of the 90\% credible region overlap.
We use the maximum \textit{a posteriori} (MAP), which is the mode of the posterior distribution, as an alternative of the 90\% confidence level to decide the credible region used for computing the overlap of the skymap between two GW signals.
We first compute the probability of the MAP of one skymap on another skymap, which is referred as the search probability $\mathcal{H}$.
It is defined as
\begin{equation}
	\mathcal{H}(p, q)=\int\mathbbm{1} \left[ q(\hat\Omega')>q(\hat\Omega_{p\textrm{, MAP}}) \right]q(\hat\Omega')d\hat\Omega',
\end{equation}
where $\mathbbm{1}$ is the indicator function, $q(\hat\Omega)$ is the posterior distribution of sky position of an event, and $\hat\Omega_{p\textrm{,MAP}}$ is the MAP estimate of $\hat\Omega_p$, also known as the coordinates of the point at highest probability density.

$\mathcal{H}(p,q)$ quantifies the overlap of the skymap between the event pairs. The lower the value of $\mathcal{H}(p,q)$, the higher the overlap between the skymaps.
The cross-HPD statistic $\mathcal{D}_{\mathrm{HPD}}(p,q)$ is defined as
\begin{equation}
	\mathcal{D}_{\mathrm{HPD}}(p,q) \equiv \max \left\{ 1-\mathcal{H}(p,q), 1-\mathcal{H}(q,p) \right\},
\end{equation}
which is symmetric in $p$ and $q$.

\section{Results and Discussion}
\label{sec:result}
We simulate 200 lensed pairs of GW signals for five different SNRs to investigate the performance of the search methodology.
We first generate the simulated data of 200 lensed pairs of GW signals with \texttt{PyCBC}~\citep{2021zndo...5347736N}.
Primary mass follows a power-law profile with slope $-2.35$, while the mass ratio and sky position are uniformly distributed.
With the simulated data, we rescale all the events to five different SNRs by adjusting the luminosity distance of each event.
In total, there are 1,000 event pairs in the injection campaign.
For details, see Sec.~\ref{res:injection}.
We use a Bayesian inference library \texttt{Bilby}~\cite{1811.02042,2006.00714} to perform parameter estimation on the GW events.
After obtaining the posterior distribution of the sky location of each event, the skymaps of all events are generated using \texttt{ligo.skymap}~\cite{1605.04242}.
We pair up the skymaps for every possible combination and separate them into lensed-pair group and nonlensed-pair group.
The three overlap statistics of all event pairs are then computed.
Finally, we plot the ROC curve to compare the performance of the overlap statistics.
Details of the ROC curve can be found in Sec.~\ref{res:performance}.

\subsection{Injection campaign}
\label{res:injection}
\begin{table}[t]
	\caption{
		The parameters of the waveform simulated and the set-up of the injection campaign.
		200 event pairs are generated.
		For simplicity, the spins of all events are set to zero.
		All events are rescaled to 5 different SNRs by adjusting the luminosity distance of each event.
		Finally, there are 2,000 events in total.
	}
	\label{table:injection}
	\centering
	\begin{ruledtabular}
		\begin{tabular}{ll}
			Number of lensed event pairs & 200                                                   \\
			Waveform approximant         & \texttt{IMRPhenomPv2}~\cite{1508.07250,1611.03703}    \\
			Spin                         & None                                                  \\
			Rescaled SNR                 & 8, 18.5, 29, 39.5, 50                                 \\
			Detector network             & HLV at design sensitivity~\cite{1408.3978,2008.01301} \\
			Primary-mass distribution    & Power law with slope $-2.35$                          \\
			Mass-ratio distribution      & Uniform                                               \\
			Sky-location distribution    & Uniform                                               \\
		\end{tabular}
	\end{ruledtabular}
\end{table}

To study the distribution of skymap overlap of lensed-pair candidates under different SNRs and the efficiency of overlap statistics on filtering potentially lensed pairs of GW signals, we set up an injection campaign with parameter settings as listed in Table \ref{table:injection}. A total of 200 lensed GW event pairs are generated.
The localization uncertainty of a GW signal depends on its $\mathrm{SNR}$, therefore we are interested in the effect of different $\mathrm{SNRs}$ on the skymaps of lensed GW signal pairs.
The uncertainty of the localization is smaller when the $\mathrm{SNR}$ is higher.
Therefore, we expect the overlap to be generally lower for GW event pairs with a higher $\mathrm{SNR}$ than those with a lower $\mathrm{SNR}$.
We use five different $\mathrm{SNR}$s to study the difference in the filtering threshold for different SNR pairs by skymap overlap.

The largest $\mathrm{SNR}$ of the event pairs is chosen to be 50, it is a sufficiently high $\mathrm{SNR}$ upper limit due to the rarity of detecting a signal with such $\mathrm{SNR}$.
After setting up the upper limit and the lower limit of the event $\mathrm{SNR}$, we uniformly divide the range between 8 and 50 to obtain five different $\mathrm{SNR}$s and rescale the luminosity distances to attain the desired SNR.

After performing the parameter estimation, we generate the skymaps for all events.
The prior of chirp mass is set to be uniform from 4 $M_{\odot}$ to 150 $M_{\odot}$ and the two component masses are constrained between 5 $M_{\odot}$ to 150 $M_{\odot}$, while the injections are drawn from an astrophysical power-law primary-mass distribution.
In LIGO-Virgo-KAGRA PE analyses, agnostic priors are typically used; these commonly assume uniform priors on the component masses~\cite{2025arXiv250818082T}.
Choosing a uniform chirp-mass prior here avoids tailoring the PE to the injections and mirrors the practical, agnostic setting in which skymap filtering would be applied.
Sky localization is dominated by network SNR, interdetector time delays and antenna patterns, so the prior mismatch does not materially affect the skymap-overlap conclusions.
The prior of geocent time is chosen to be a delta function that peaks at the injected parameter. The sampler used for the parameter estimation is \texttt{dynesty}~\cite{1904.02180}.
Finally, the three overlap statistics of each event pair are computed.
The lensed event pairs consist of 1000 samples which are 200 lensed events pairs for each of the five SNRs.
For the nonlensed event pairs, the overlap of all the possible event pairs from the original 2,000 events is computed, which results in about 700,000 samples.
By plotting the ROC curve of each statistic, we study the performance of the three statistics on filtering lensed candidate event pairs.

\subsection{Coverage of samples}
\label{res:coverage}
\begin{table}[t]
	\caption{
		Mean area of the 90\% credible region for skymaps from 200 simulated events using the HLV network at five SNRs.
		Details of the injection campaign are given in Table~\ref{table:injection}.
	}
	\label{table:coverage}
	\centering
	\begin{ruledtabular}
		\begin{tabular}{S[table-format=2.1] S[table-format=3.1]}
			{SNR} & {Mean area of 90\% credible region ($\mathrm{deg}^2$)} \\ \hline
			8     & 186.8                                                  \\
			18.5  & 34.8                                                   \\
			29    & 12.0                                                   \\
			39.5  & 6.4                                                    \\
			50    & 4.1
		\end{tabular}
	\end{ruledtabular}
\end{table}

\begin{figure*}
	\centering
	\includegraphics[width=\linewidth]{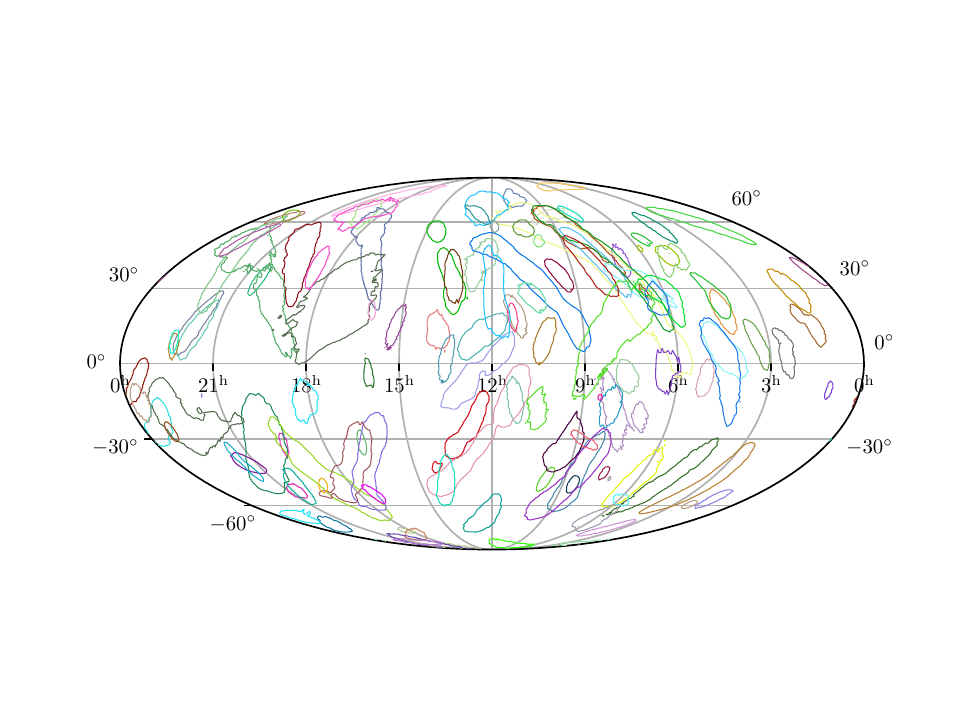}
	\caption{Contours of 90\% credible region of 200 skymaps with $\mathrm{SNR}=8$ are shown in the figure.
		Details of the injection campaign are referred to Table~\ref{table:injection}.
		The figure shows the sky coverage of the injections with $\mathrm{SNR} = 8$.}
	\label{fig:coverage}
\end{figure*}

Table \ref{table:coverage} shows the mean area of 90\% credible region of the skymaps in the injection campaign.
The size of the 90\% credible region relates to the uncertainty of the localization.
For GW signals with a smaller $\mathrm{SNR}$, the uncertainty of localization is larger.
Therefore, the mean area of 90\% credible region for events with $\mathrm{SNR}=8$ is significantly larger than that of the events with $\mathrm{SNR}=50$.

Figure \ref{fig:coverage} illustrates the 90\% credible region of events with $\mathrm{SNR}=8$.
The contours of their 90\% credible regions are demonstrated in different colors for distinguishing different contours.
The 200 events come from different parts of the sky and their 90\% credible regions covered about half of the sky.
Note that there are overlaps of 90\% credible regions of nonimage pairs.
Those nonimage event pairs with relatively high overlap values are the false positives.
The coverage ensures that there is a sufficient number of lensed event pairs in the injection campaign to study the distribution of skymap overlap for nonimage pairs.

\subsection{Performance of statistics}
\label{res:performance}
With the overlap statistics of all the event pairs, the performance of the statistics is shown in a ROC curve.
In a ROC curve, which shows the performance of a classification model at all classification thresholds by plotting the true positive rate (TPR) against the FPR.

The y axis of the ROC curves is TPR, which is defined as
\begin{equation}
	\mathrm{TPR} = \frac{\mathrm{TP}}{\mathrm{TP}+\mathrm{FN}},
\end{equation}
where TP is the number of lensed event pairs which overlap value is higher than the filtering threshold, while FN is the number of lensed event pairs which overlap values are smaller than the filtering threshold.

The x axis of the ROC curves is FPR, which is defined as
\begin{equation}
	\mathrm{FPR} = \frac{\mathrm{FP}}{\mathrm{FP}+\mathrm{TN}},
\end{equation}
where FP is the number of nonlensed event pairs which overlap value is higher than the filtering threshold, while TN is the number of nonlensed event pairs which overlap value is smaller than the filtering threshold.

\begin{figure*}
	\centering
	\includegraphics[width=0.8\textwidth]{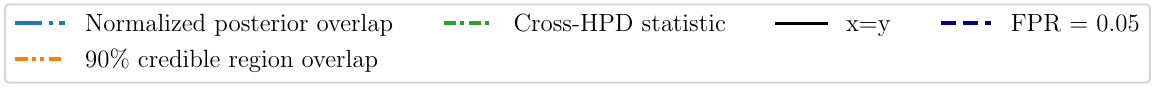}
	\centering
	\subfigure[$\mathrm{SNR}=8$ vs $\mathrm{SNR}=8$]{
		\includegraphics[width=0.47\textwidth]{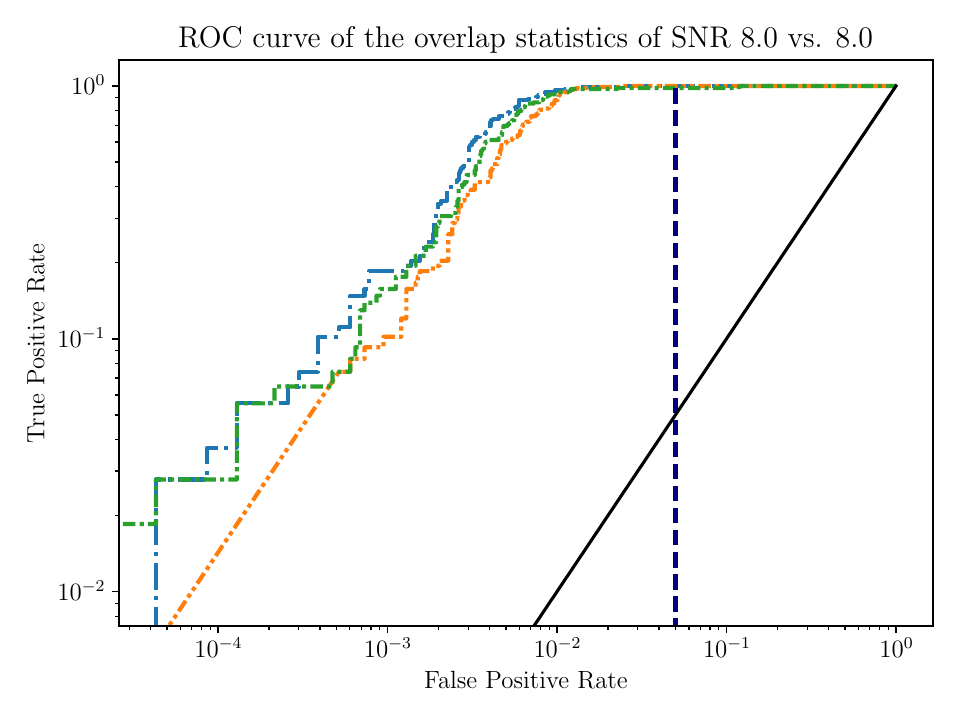}
		\label{fig:roc_(8.0_8.0)}
	}
	\subfigure[$\mathrm{SNR}=8$ vs $\mathrm{SNR}=18.5$]{
		\includegraphics[width=0.47\textwidth]{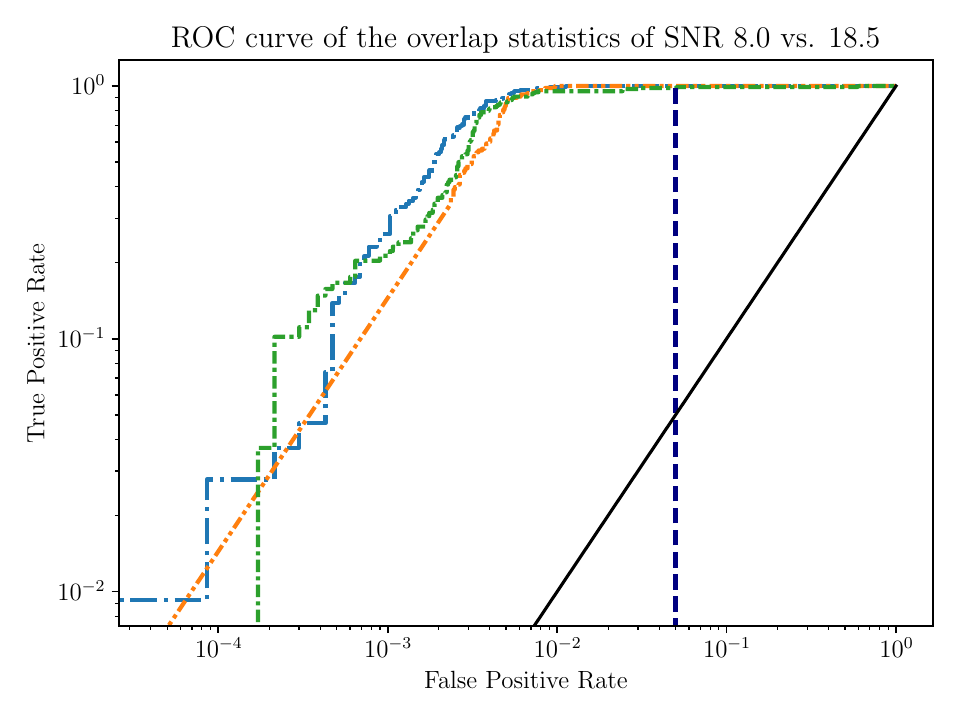}
		\label{fig:roc_(8.0_18.5)}
	}
	\subfigure[$\mathrm{SNR}=8$ vs $\mathrm{SNR}=29$]{
		\includegraphics[width=0.47\textwidth]{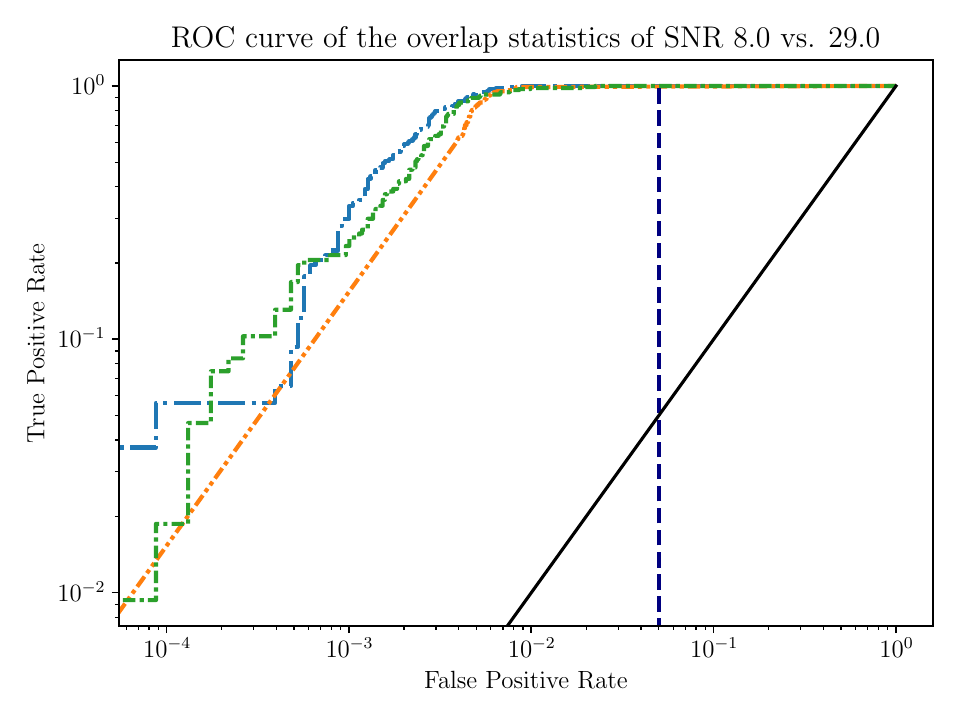}
		\label{fig:roc_(8.0_29.0)}
	}
	\subfigure[$\mathrm{SNR}=8$ vs $\mathrm{SNR}=39.5$]{
		\includegraphics[width=0.47\textwidth]{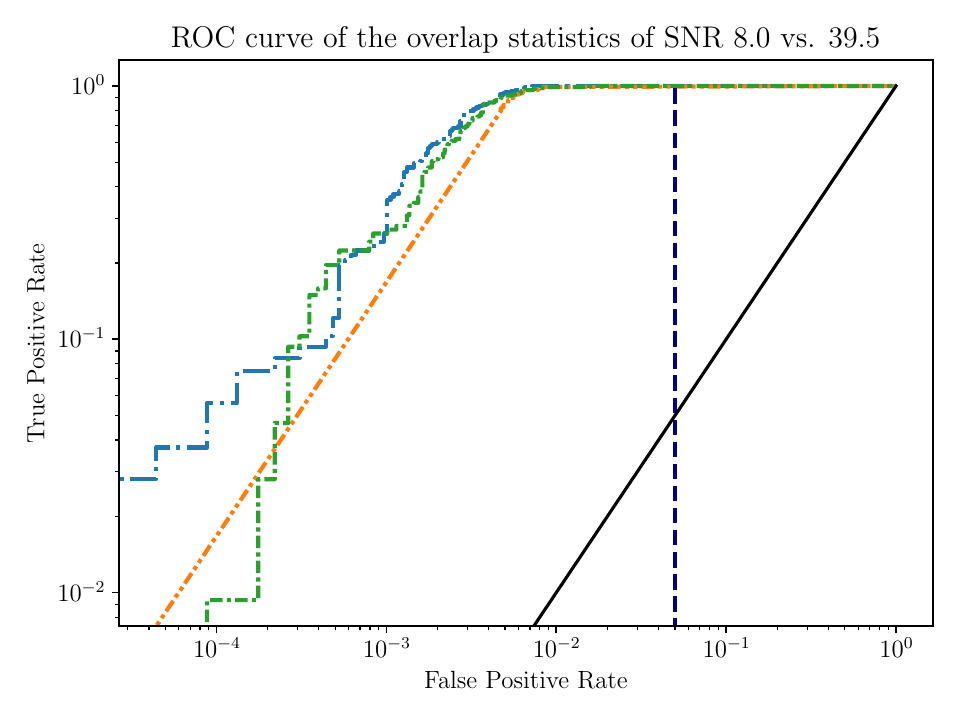}
		\label{fig:roc_(8.0_39.5)}
	}
	\subfigure[$\mathrm{SNR}=8$ vs $\mathrm{SNR}=50$]{
		\includegraphics[width=0.47\textwidth]{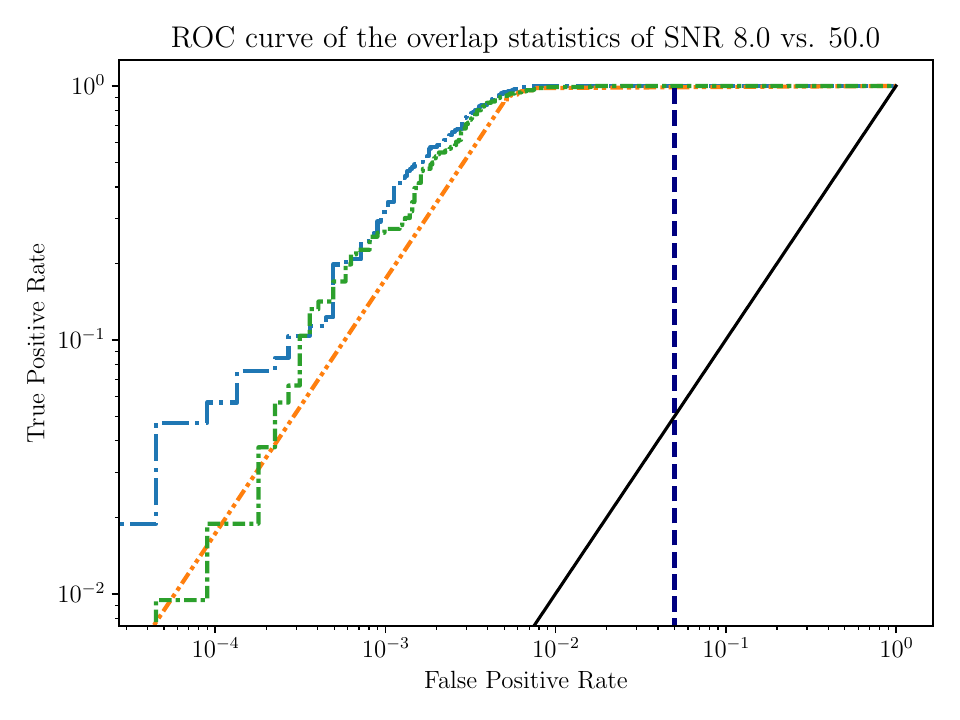}
		\label{fig:roc_(8.0_50.0a)}
	}


	\caption{ROC curves of event pairs with one event $\mathrm{SNR}=8$.
		The black solid line $x=y$ represents the ROC curve of a random classifier.
		ROC curves indicate the performance of the overlap statistics.
		The classifier is better if it can achieve a higher true positive rate given a false positive rate.
		The three statistics do not show a large difference in the region of FPR between $10^{-2}$ and 1.}
	\label{fig:low_snr_roc}
\end{figure*}

\begin{figure*}
	\centering
	\includegraphics[width=0.8\textwidth]{legend.pdf}
	\subfigure[SNR=8 vs SNR=50]{
		\includegraphics[width=0.47\textwidth]{roc_8.0_50.0.pdf}
		\label{fig:roc_(8.0_50.0)}
	}
	\subfigure[SNR=18.5 vs SNR=50]{
		\includegraphics[width=0.47\textwidth]{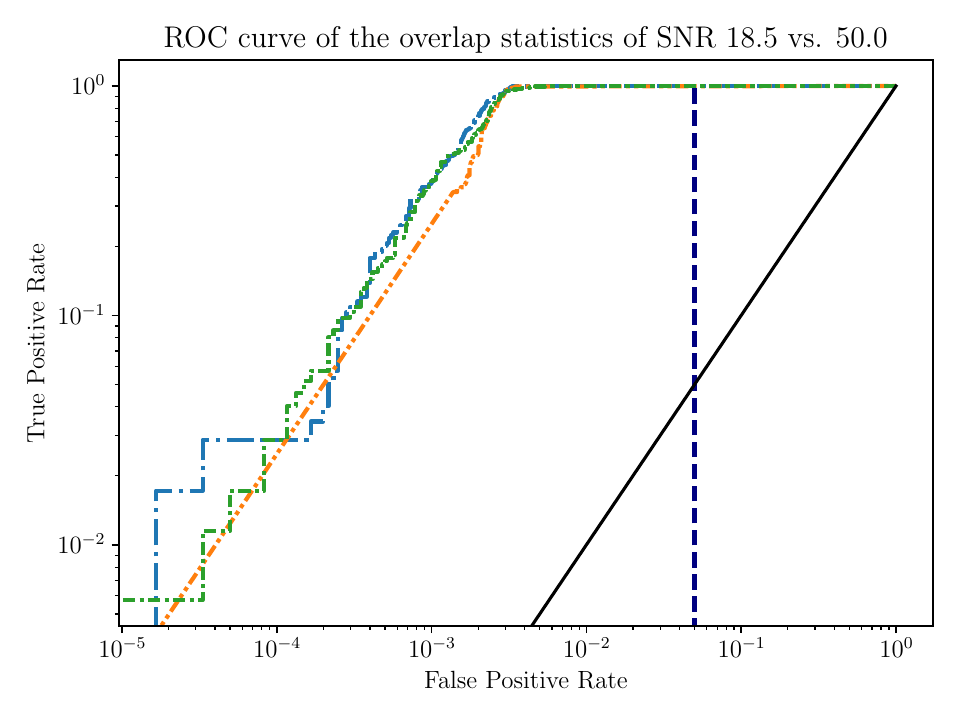}
		\label{fig:roc_(18.5_50.0)}
	}
	\subfigure[SNR=29 vs SNR=50]{
		\includegraphics[width=0.47\textwidth]{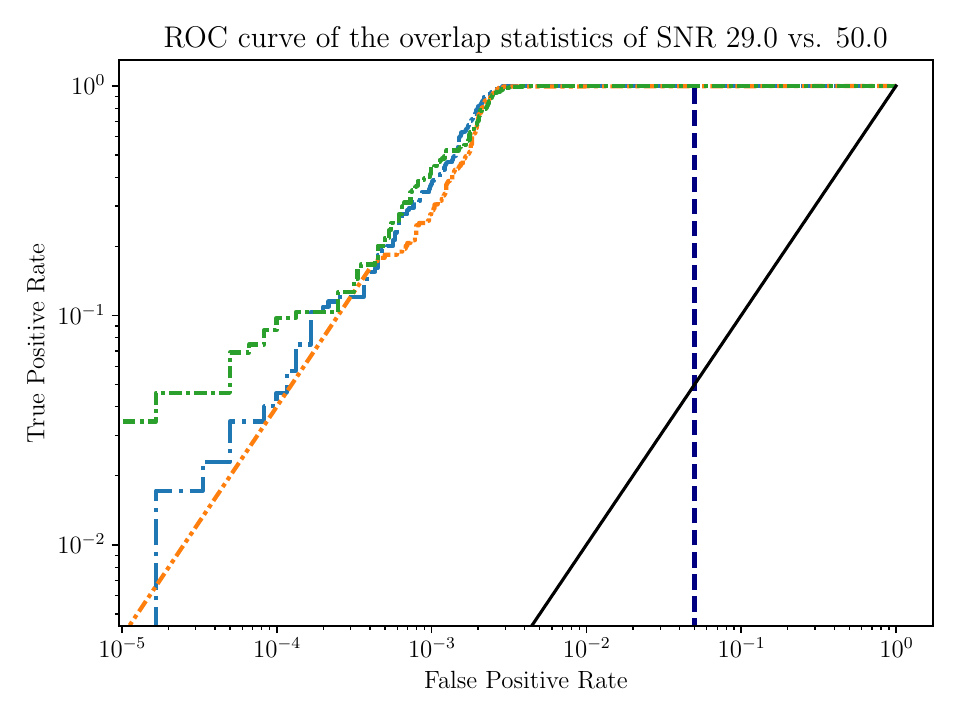}
		\label{fig:roc_(29.0_50.0)}
	}
	\subfigure[SNR=39.5 vs SNR=50]{
		\includegraphics[width=0.47\textwidth]{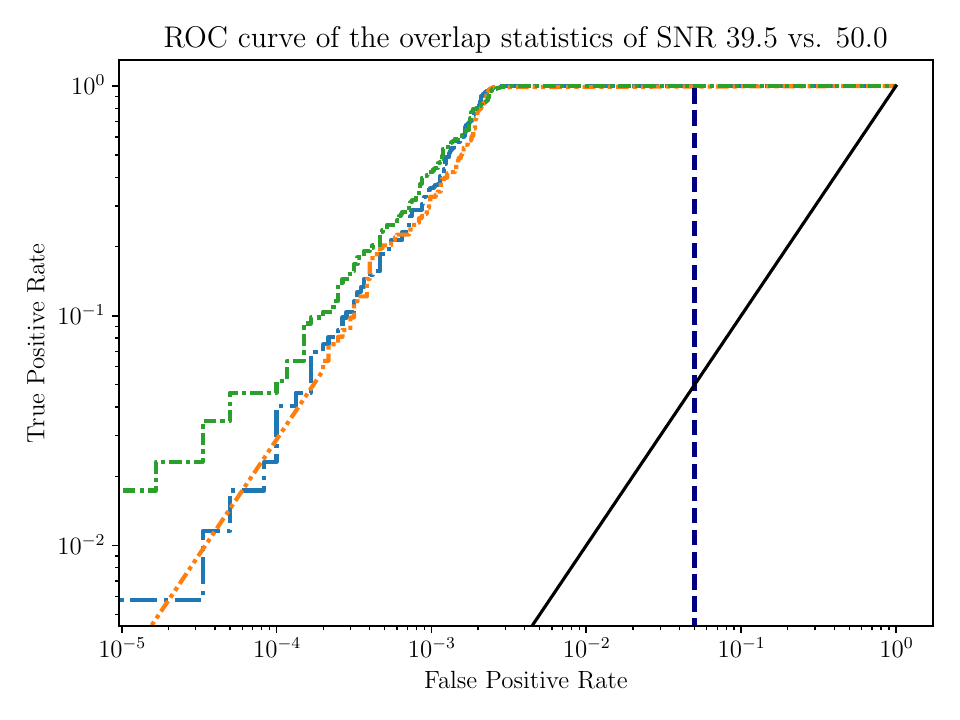}
		\label{fig:roc_(39.5_50.0)}
	}
	\subfigure[SNR=50 vs SNR=50]{
		\includegraphics[width=0.47\textwidth]{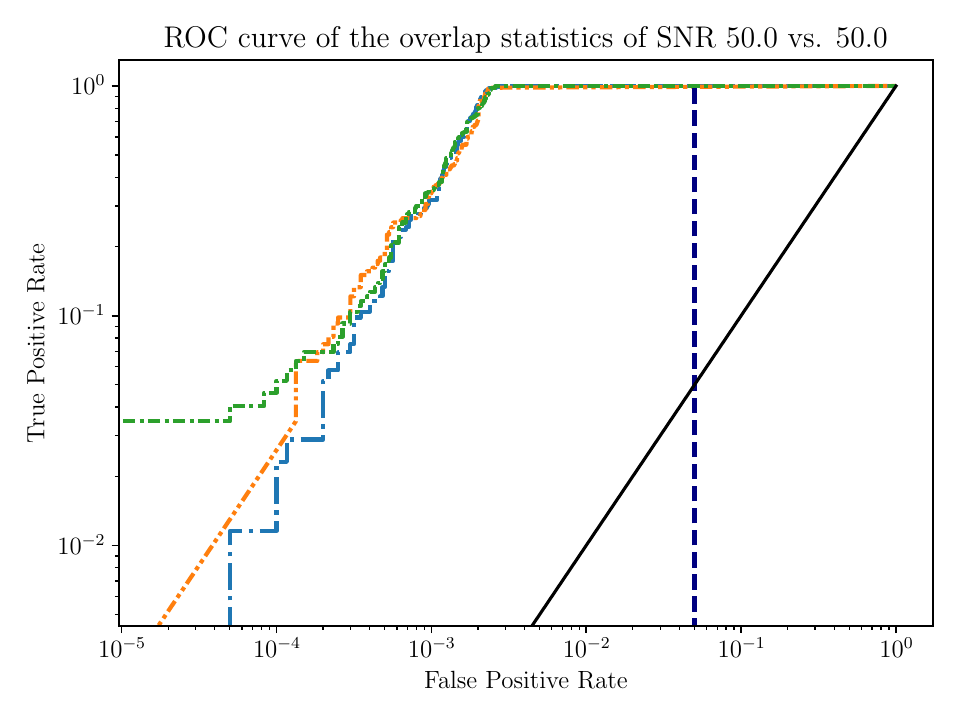}
		\label{fig:roc_(50.0_50.0)}
	}
	\caption{ROC curves of event pairs with one event $\mathrm{SNR}=50$.
		The black solid line $x=y$ represents the ROC curve of a random classifier.
		The classifier is better if it can achieve a higher true positive rate given a false positive rate.
		The three statistics do not show a large difference in the region of FPR between $10^{-2}$ and 1.}
	\label{fig:high_snr_roc}
\end{figure*}

Given a specific FPR or a specific filtering threshold, a higher TPR indicates a better performance of the statistic because less lensed event pairs are recognized as nonlensed event pairs by the statistic.
We aim to set up a reasonable threshold with $\mathrm{TPR}\approx1$ to filter nonlensed event pairs while keeping most of the lensed event pairs for further analysis.
Ten ROC curves are shown in Fig. \ref{fig:low_snr_roc} and Fig. \ref{fig:high_snr_roc}.
The five ROC curves in Fig. \ref{fig:low_snr_roc} indicate the performance of the three statistics for event pairs with one of the event SNRs being small ($\mathrm{SNR}=8$), while the five ROC curves in Figure \ref{fig:high_snr_roc} indicate the performance of the three statistics for event pairs with one of the event SNRs is large ($\mathrm{SNR}=50$).


Figure~\ref{fig:low_snr_roc} shows the ROC curves of the three overlap statistics for event pairs with one event SNR$=8$ while another is one of the five SNRs.
In this case, the TPR of the three statistics shows nearly no difference when the FPR is fixed to a value between $10^{-2}$ and 1.
The difference in the performance of the statistics is noticeable only when we fix FPR to a value smaller than around $10^{-2}$.

For a relatively small FPR value ($\mathrm{{FPR}<10^{-2}}$), 90\% credible region overlap shows the worst overall performance as it shows the most significant drop at a lower FPR.
However, all three statistics show a significant drop when the FPR is lower than $10^{-2}$. Within this range, the losses of lensed event pairs are significant for all three statistics.
Therefore, the threshold should be fixed higher than $10^{-2}$ if we aim to keep most of the lensed event pairs.
Thus, the performance of the three statistics is similar for a reasonable choice of threshold that can filter out nonlensed event pairs while keeping most of the lensed event pairs.
The straight line of the TPR drop for 90\% credible region overlap in $\mathrm{FPR} \lessapprox 10^{-3}$ is due to a lack of samples.
The data points of the ROC curve are generated by using a threshold to separate the input data into FP, TP, FN, and TN.

Figure \ref{fig:high_snr_roc} shows the ROC curves of the three overlap statistics for event pairs with one event $\mathrm{{SNR}=50}$ while another is one of the five SNRs.
In this case, the TPR of the three statistics shows nearly no difference when the FPR is fixed to a value between $10^{-2}$ and 1.
The difference in the performance of the statistics is noticeable only when we fix FPR to a value smaller than around $10^{-3}$.

For a relatively small FPR value ($\mathrm{{FPR}<10^{-3}}$), 90\% credible region overlap shows the worst overall performance.
However, again, choosing a threshold with FPR lower than around $10^{-3}$ will lead to a significant loss in the number of lensed event pairs.
Therefore, the filtering threshold should be fixed at an FPR higher than $10^{-3}$ in order to keep all of the lensed event pairs.
Thus, the performance of the three statistics is similar for a reasonable choice of filtering threshold that can filter out nonlensed event pairs while keeping all of the lensed event pairs.
Again, the straight line of the TPR drop for 90\% credible region overlap in $\mathrm{{FPR} \lessapprox 10^{-3}}$ is due to the lack of samples.

For the purpose of filtering potentially lensed candidate event pairs, the three skymap overlap statistics demonstrated a similar performance while fixing FPR between $10^{-2}$ and 1.
In the mock data study, we can set up a specific $\mathrm{FPR}=10^{-2}$ for event pairs in all SNR to filter out 99\% of the nonlensed candidate pairs, which demonstrated that a large fraction of nonimage pairs could be rapidly filtered away using the skymap overlap.

\section{Conclusions}
\label{sec:conclusion}

In this paper, we have demonstrated our proposed search methodology using skymap overlap to filter potentially lensed pairs of GW signals.
By setting up a threshold with a specific $\mathrm{FPR}=10^{-2}$ for the three overlap statistics, we can filter out more than 99\% of the nonlensed event pairs while keeping most of the lensed event pairs, which shows the feasibility of using skymap overlap to filter potentially lensed event pairs of GW signals.

Since the computation of the three statistics for each event pair takes less than 3 seconds, a new trigger could be constructed to compute the skymap overlap of all existing events within a short period of time using the low latency skymaps generated by \texttt{BAYESTAR}~\cite{1508.03634} or other rapid sky localization algorithms.

To accommodate real detections, which span continuous SNR combinations, we propose generating a dense injection grid to derive ROC-based thresholds as a function of SNR and interpolating these results to produce a runtime mapping that adjusts the overlap threshold for any observed pair; this enables real-time, SNR-conditioned filtering in production pipelines.


\section*{Acknowledgements}
\label{sec:acknowledgements}

We thank Otto Akseli Hannuksela for fruitful discussion.
I. C. F. W. and T. G. F. L. are partially supported by grants from the Research Grants Council of the Hong Kong (Projects No. 24304317 and No. 14306419) and Research Committee of the Chinese University of Hong Kong.
R. K. L. L. and T. G. F. L. would also like to gratefully acknowledge the support from the Croucher Foundation in Hong Kong.
The authors acknowledge the use of the IUCAA LDG cluster Sarathi for the computational/numerical work.
The authors are also grateful for the computational resources provided by the Chinese University of Hong Kong.
The Center of Gravity is a Center of Excellence funded by the Danish National Research Foundation under Grant No. 184.
This work is also partially supported by the Research Foundation - Flanders (FWO) through Grants No. I002123N, No. I000725N, No. G086722N.

\section*{Data availability}
\label{sec:data_availability}

The data that support the findings of this manuscript are openly available \cite{chan_2025_17177201}.

\bibliographystyle{apsrev}
\bibliography{references}

\end{document}